\documentclass{raa}
\usepackage{latexsym}
\usepackage{dcolumn}
\usepackage{bm}
\usepackage{comment}
\usepackage{color}
\usepackage{epsfig}
\usepackage{mathrsfs}
\usepackage{hyperref}
 
\usepackage{xcolor}
\usepackage{mathtools}
\usepackage{relsize}
\usepackage{booktabs} 

\usepackage{graphicx,times}
\usepackage{natbib}
\usepackage{amssymb,amsmath}
\usepackage{longtable}
\bibpunct{(}{)}{;}{a}{}{,} 

\begin{document}
\title{Cosmological constraints on ultra-light axion fields}
%\volnopage{Vol.000,No. 000--000} 
%\setcounter{page}{1}

\author{Jiangang Kang\inst{1,2}, Yan Gong\inst{*1}, Gong Cheng\inst{1,2}, Xuelei Chen\inst{1,2,3}}
\institute{Key Laboratory for Computational Astrophysics, National Astronomical Observatories, Chinese Academy of Sciences, 20A Datun Road, Beijing 100101, China\\
\and
School of Astronomy and Space Science,University of Chinese Academy of Sciences, Beijing 100049, China
\and
Centre for High Energy Physics, Peking University, Beijing 100871, China
}
\date{\today}
\abstract {Ultra-light axions (ULAs) with mass less than $10^{-20}$ eV have interesting behaviors that may contribute to either dark energy or dark matter at different epochs of the Universe. Its properties can be explored by cosmological observations, such as expansion history of the Universe, cosmic large-scale structure, cosmic microwave background, etc. In this work, we study the ULAs with a mass around $10^{-33}$ eV, that means the ULA field still rolls slowly at present with the equation of state $w=-1$ as dark energy. In order to investigate the mass and other properties of this kind of ULA field, we adopt the measurements of Type Ia supernova (SN Ia), baryon acoustic oscillation (BAO), and Hubble parameter $H(z)$. The Markov Chain Monte Carlo (MCMC) technique is employed to perform the constraints on the parameters. Finally, by exploring four cases of the model, we find that the mass of this ULA field is about $3\times 10^{-33}$ eV if assuming the initial axion field $\phi_{\rm i}=M_{\rm pl}$. We also investigate a general case by assuming $\phi_{\rm i}\le M_{\rm pl}$, and find that the fitting results of $\phi_{\rm i}/M_{\rm pl}$ are consistent with or close to 1 for the datasets we use.}
\maketitle

\section{Introduction}\label{intro}

Axion or axion-like particle (ALP) is a good candidate for cold dark matter (CDM) \citep{1977PhRvL..38.1440P,1978PhRvL..40..223W,1978PhRvL..40..279W}. It was proposed to solve the strong CP problem in the Quantum Chromodynamics (QCD) theory, and generated by breaking the Peccei-Quinn (PQ) symmetry \citep{1977PhRvL..38.1440P}. Axion has a huge possible mass range spanning over many orders of magnitude, that cannot be stringently constrained by theory. In the string theory, axion is allowed to have extremely small mass, i.e. ultra-light axion (ULA), which ranges from $10^{-20}$ to $10^{-33}$ eV or even smaller \citep{1984CMaPh..92..455W, 2006JHEP...06..051S,2016PhR...643....1M}. Some interesting properties of ULAs appear in this mass range. 

In the  early Universe, the Hubble parameter $H$ is larger than axion mass $m_a$, i.e. $H>m_a$ (the natural units are used with $\hbar=1$ hereafter). Then the axion field is overdamped by the Hubble friction and would roll slowly. It means the ULA equation of state $w_a=-1$ acting as dark energy with negative pressure. Later, when the Universe expanding slowly and slowly, we have $H<m_a$. At this time, the ULA field is underdamped and begins to oscillate. The equation of state of ULA also oscillates correspondingly between -1 and 1, which has an average value $w\simeq0$ acting as dark matter. Since the mass of ULAs is quite small, the de Broglie wavelength of ULAs can be large enough to affect the formation of cosmic structure at galaxy or sub-galaxy scales, which is similar to the free-streaming effect of warm dark matter. Therefore, the ULAs can contribute to both dark energy and dark matter at different epochs of the Universe.

For $10^{-20}<m_a<10^{-27}$, the ULAs begin to oscillate before the epoch of the radiation-matter equality ($H(z_{\rm eq})\sim 10^{-28}$ eV), and since then behave like dark matter with energy density $\rho_a\sim a^{-3}$, that can contribute to all of dark matter in this mass range\citep{2017PhRvL.119c1302I,2017MNRAS.471.4606A,Bar_2019,PhysRevLett.123.051103,Nebrin_2019}. On the other hand, for $m_a<10^{-27}$ eV, the ULAs will oscillate after the radiation-matter equality, and can contribute to the late accelerating expansion of the Universe as dark energy \citep{2010PhRvD..82j3528M, 2015PhRvD..91j3512H,2016PhR...643....1M}.

In this work, we explore the ULAs with mass $m_a  \sim 10^{-33}$, which means $m_a\sim H_0$ where $H_0\simeq1.5\times10^{-33}$ eV is the present Hubble constant. It implies that the axion field has not or just entered the oscillation stage and still acts as dark energy at present with the equation of state around -1. The data of Type Ia supernova (SN Ia), baryon acoustic oscillation (BAO), and Hubble parameters at different redshifts $H(z)$ are used to constrain this kind of ULA filed and its mass. The Markov Chain Monte Carlo (MCMC) method is adopted in the constraint to derive the probability distribution functions (PDFs) of the parameters in the model. 

The paper is organized as follows: in Section 2, we discuss the ultra-light axion model and derive the relevant equations used in the constraint; in Section 3, we describe the SN Ia, BAO, and H(z) data we adopt in the fitting process; in Section 4, we show the constraint result; we summarize and discuss the result in Section 5.

\section{Model}\label{model}

The action of the ultra-light axion field is given by
\begin{equation}
%S = \int dx^4 \sqrt{-g}\left[  -\frac{1}{2}f_a^2(\partial \theta)^2-\Lambda_a^4P(\theta)  \right],
S_{\phi} = \int dx^4 \sqrt{-g}\left[  -\frac{1}{2}(\partial \phi)^2-V(\phi)  \right],
\end{equation}
where $\phi$ is the canonically normalized axion field which has a shift symmetry $\phi\to \phi+{\rm const}$, and $V(\phi)$ is a periodic potential with the minimum at $\phi=0$. A simple choice of $V(\phi)$ is given by \citep{2010PhRvD..82j3528M, 2015PhRvD..91j3512H,2016PhR...643....1M}.
\begin{equation}
V(\phi) = \Lambda_{a}^4 \left( 1-{\rm cos}\frac{\phi}{f_a} \right).
\end{equation}
Here $\Lambda_{a}$ is the amplitude of the potential which indicates the energy scale of non-perturbative physics, and $f_a$ is the PQ symmetry-breaking energy scale. Expanding the potential for $\phi\ll f_a$ as a Taylor series, then the dominant term is
\begin{equation} \label{eq:Va}
V(\phi) \simeq \frac{1}{2}m_a^2 \phi^2,
\end{equation}
where the axion mass is given by $m_a^2=\Lambda_a^4/f_a^2$. We can find that the axion mass can be extremely small if $f_a\gg\Lambda_a$, which is the case we discuss in this work. 

Assuming a flat Friedmann-Robertson-Walker (FRW) metric, the variation of the action with $V(\phi)$ given in Eq.~(\ref{eq:Va}), the equation of motion can be derived as
\begin{equation} \label{eq:eom}
\ddot{\phi} + 3H\dot{\phi} + m_a^2\phi = 0.
\end{equation}
Here $H=\dot{a}/a$ is the Hubble parameter, and $a$ is the scale factor. The  energy density and pressure of the axion field can be also obtained from the energy momentum tensor as
\begin{eqnarray} \label{eq:rho}
\rho_a &=& \frac{1}{2}\dot{\phi}^2 + \frac{1}{2}m_a^2\phi^2, \\
P_a &=& \frac{1}{2}\dot{\phi}^2 - \frac{1}{2}m_a^2\phi^2.\label{eq:P}
\end{eqnarray}

In the radiation or matter dominated era, we have $a(t)\propto t^p$ where $p=1/2$ or $2/3$, respectively. In this case, Eq.~(\ref{eq:eom}) has an exact solution, and it is found that $\phi={\rm const}$ for $m_at\ll 1$ at early time ($a<a_{\rm osc}$), and $\phi$ begins to oscillate for $m_at\gg 1$ at late time ($a>a_{\rm osc}$), where $a_{\rm osc}$ is the scale factor when oscillation occurs \citep{2015PhRvD..91j3512H,2016PhR...643....1M}. Then the equation of state of the axion field $w=P_a/\rho_a$ can be derived from Eq.~(\ref{eq:rho}) and (\ref{eq:P}). For extremely small $m_a$ (or $m_at\ll 1$), $\phi=\rm const$ and $\dot{\phi}=0$, and we can find that $w=-1$, which means the axion filed behaves like dark energy (DE, e.g. cosmological constant) with constant energy density. On the other hand, for large $m_a$ (or $m_at\gg 1$), $w$ will oscillate around 0 between -1 and 1, that acts like ordinary matter.
 
For DE-like axion field, the energy density parameter $\Omega_a=\rho_a/\rho_{c0}$, where $\rho_a=1/2\,m_a^2\phi^2$, and $\rho_{c0}=3H_0^2/8\pi G=3H_0^2M_{\rm pl}^2$ is the current critical density, where $H_0$ is the Hubble constant in $\rm eV$ and $M_{\rm pl}$ is the Planck mass. Considering $m_a$ is quite small that the ultra-light axion field has not begun to oscillate at present (i.e. $a_{\rm osc}>1$), and then we have
\begin{equation} \label{eq:Omega_a}
\Omega_a = \frac{1}{6} \left(\frac{m_a}{H_0}\right)^2 \left( \frac{\phi_{\rm i}}{M_{\rm pl}} \right)^2,
\end{equation}
where $\phi_{\rm i}=\phi(a<a_{\rm osc})$ is the initial homogeneous axion field, and $f_{\phi}= \phi_{\rm i}/M_{\rm pl}\lesssim 1$ \citep{2015PhRvD..91j3512H,2016PhR...643....1M}. We will explore the constraints for $f_{\phi}=1$ and $f_{\phi}\le1$ cases, respectively,  in the following discussion. Then, in our model, the Hubble parameter can be written as
\begin{eqnarray}
H(z) &=& H_0 \left[ (1-\Omega_a)(1+z)^3 + \Omega_a\right]^{1/2}\ \ {\rm in\ flat\ case},\\
H(z) &=& H_0 \left[\Omega_m(1+z)^3 + \Omega_a + \Omega_k(1+z)^2\right]^{1/2}\ \ {\rm in\ nonflat\ case} .
\end{eqnarray}
Here $H_0=100\, h\ \rm km\,s^{-1}Mpc^{-1}$, $\Omega_m$ is the matter energy density parameter, and $\Omega_k=1-\Omega_m-\Omega_a$ is the cosmic curvature parameter. The comoving distance can be estimated by
\begin{equation}
D_{\rm C}(z) = \frac{1}{\sqrt{|\Omega_k|}} {\rm sinn}\left[\sqrt{|\Omega_k|}\int_0^z\frac{\ c\,dz'}{H(z')}\right],
\end{equation}
where ${\rm sinn}(x)={\rm sinh}(x)$, $x$, and ${\rm sin}(x)$ for open, flat, and closed geometries, respectively.

\section{Data}\label{data}

\subsection{SN Ia data}

We adopt Pantheon SN Ia sample in the constraints, which contains 1048 SNe Ia from Pan-STARRS1 (PS1), Sloan Digital Sky Survey (SDSS),  SNLS, and various low-z and Hubble Space Telescope samples in the range $0.01<z<2.3$ \citep{2018ApJ...859..101S}. The $\chi^2$ distribution is used to estimate the likelihood function $\mathcal{L}\propto {\rm exp}(-\chi^2/2)$, and it can be written as
\begin{equation}
\chi^2_{\rm SN}=\Delta {\bf m}^T\cdot{\bf C_{\rm m}}^{-1}\cdot\Delta {\bf m}.
\end{equation}
Here $\Delta{\bf m}={\bf m}_{\rm obs}-{\bf m}_{\rm th}$ where ${\bf m}_{\rm obs}$ and ${\bf m}_{\rm th}$ are the vectors of observational and theoretical apparent magnitudes, respectively, and ${\bf C_{\rm m}}$ is the covariance matrix. $m_{\rm th}=\mu_{\rm th}+M$ where $\mu_{\rm th}$ is the theoretical distance modulus, and $M$ is the absolute magnitude. It can be further expressed as
\begin{eqnarray}
m_{\rm th} &=& 5{\rm log}_{10}D_{\rm L}(z)+25+M, \label{eq:mu_M}\\
                  &=& 5{\rm log}_{10}{\mathcal D}_{\rm L}(z)+{\mathcal M}.
\end{eqnarray}
Here $D_{\rm L}(z)=(1+z)D_{\rm C}(z)$ is the luminosity distance, ${\mathcal D}_{\rm L}$ is the Hubble-constant free luminosity distance, and ${\mathcal M}$ is a nuisance parameter that is combined with the Hubble constant and $M$. The covariance matrix $\bf C$ takes the form as
\begin{equation}
{\bf C} = {\bf D}_{\rm stat} + {\bf C}_{\rm sys},
\end{equation}
where ${\bf D}_{\rm stat}$ is the vector of statistic error, which includes photometric error of the SN distance, uncertainties of distance from the mass step correction, the distance bias correction, the peculiar velocity, redshift measurement, stochastic gravitational lensing, and the intrinsic scatter. ${\bf C}_{\rm sys}$ is the systematic covariance matrix for the data \citep{2018ApJ...859..101S}.

\subsection{BAO data}

\begin{table*}[ht]
	\vspace{1mm} 
	\begin{center}
	\caption{The BAO data used in this work. There are 16 data points from $z=0.1$ to 2.4 are included.} \label{tab:BAO_data}
		\begin{tabular}{c|c|c|c|c|c}
			\hline  \hline
			Redshift &Measurement& Value &$r_{\rm s, fid}$&Survey & References\\[0.8ex]
			\hline
			0.106 &$r_{\rm s}/D_{\rm V}$& 0.3360$\pm$0.015 &-- &6dFGS  &\citep{2011MNRAS.416.3017B}\\
			0.15 &$r_{\rm s}/D_{\rm V}$&0.2239$\pm$0.0084  &-- &SDSS DR7 &\citep{2015MNRAS.449..835R}\\
			0.32 &$r_{\rm s}/D_{\rm V}$&0.1181$\pm$0.0024  &-- &BOSS LOW-Z &\citep{2014MNRAS.441...24A}\\
			0.57 &$r_{\rm s}/D_{\rm V}$&0.0726$\pm$0.0007  &-- &BOSS CMASS &\citep{2012MNRAS.427.2132P}\\
			0.44 &$r_{\rm s}/D_{\rm V}$&0.0870$\pm$0.0042 &-- &WiggleZ &\citep{Blake2012}\\
			0.60 &$r_{\rm s}/D_{\rm V}$&0.0672$\pm$0.0031 &-- &WiggleZ &\citep{Blake2012}\\
			0.73 &$r_{\rm s}/D_{\rm V}$&0.0593$\pm$0.0020 &-- &WiggleZ &\citep{Blake2012}\\
			2.34 &$r_{\rm s}/D_{\rm V}$&0.0320$\pm$0.0013  &-- &SDSS-III DR11 &\citep{2015AA...574A..59D}\\
			2.36 &$r_{\rm s}/D_{\rm V}$&0.0329$\pm$0.0009  &-- &SDSS-III DR11 &\citep{2015JPhD...48R3001H}\\
			0.15 &$D_{\rm V}(r_{\rm s, fid}/r_{\rm s})$&664$\pm$25  &148.69 &SDSS DR7 &\citep{2015PhRvD..92l3516A}\\
			1.52 &$D_{\rm V}(r_{\rm s, fid}/r_{\rm s})$&3843$\pm$147  &147.78 &SDSS DR14 &\citep{2018MNRAS.473.4773A}\\
			0.38 &$D_{\rm M}(r_{\rm s, fid}/r_{\rm s})$&1518$\pm$22  &147.78 &SDSS DR12 &\citep{2017MNRAS.470.2617A}\\
			0.51 &$D_{\rm M}(r_{\rm s, fid}/r_{\rm s})$&1977$\pm$27  &147.78 &SDSS DR12 &\citep{2017MNRAS.470.2617A}\\
			0.61 &$D_{\rm M}(r_{\rm s, fid}/r_{\rm s})$&2283$\pm$32 &147.78 &SDSS DR12 &\citep{2017MNRAS.470.2617A}\\
			2.40 &$D_{\rm M}/r_{\rm s}$&36.6$\pm$1.2  &--        &SDSS DR12   &\citep{2017AA...608A.130D}\\
			2.40 &$D_{\rm H}/r_{\rm s}$&8.94$\pm$0.22 &--        &SDSS DR12    &\citep{2017AA...608A.130D}\\
			\hline
		\end{tabular}
	\end{center}
	\vspace{-2mm}
\end{table*}

In Table~{\ref{tab:BAO_data}}, we list the adopted quantities that derived from the BAO surveys. $r_{\rm s}$ is the radius of the comoving sound horizon at the drag epoch, which takes the form as
\begin{equation}
r_{\rm s} = \int_0^{t_{\rm s}}c_{\rm s}\frac{{\rm d}t}{a},
\end{equation}
where $c_{\rm s}$ is the sound speed, $t_{\rm s}$ is the epoch of last scattering, $a$ is the scale factor. Since $r_{\rm s}$ are not sensitive to physics at low redshifts, we find that it will not be affected in our model significantly. Hence, for simplicity, we fix the value as $r_{\rm s}=147.09\pm0.26$ given by {\it Planck}~2018 result \citep{2018arXiv180706209P}. The spherically averaged distance $D_{\rm V}$ is given by
\begin{equation}
D_{\rm V}(z) = \left[ D_{\rm M}^2(z)\frac{cz}{H(z)} \right]^{1/3},
\end{equation}
where $D_{\rm M}(z)=(1+z)D_{\rm A}$ is the comoving angular diameter distance, and $D_{\rm A}=D_{\rm C}/(1+z)$ is the physical angular diameter distance. $D_{\rm H}=c/H(z)$ is the Hubble distance.

The $\chi^2$ for the BAO data can be estimated by
\begin{equation}
\chi^2_{\rm BAO} =\Delta {\bf D}^T\cdot{\bf C_{\rm D}}^{-1}\cdot\Delta {\bf D},
\end{equation}
where ${\bf D}={\bf D}_{\rm obs}-{\bf D}_{\rm th}$, and ${\bf D}_{\rm obs}$ and ${\bf D}_{\rm th}$ are the observational and theoretical quantities shown in Table~\ref{tab:BAO_data}. $\bf C_{\rm D}$ is the corresponding covariance matrix.

\subsection{$H(z)$ data}

The $H(z)$ data are also used in this work, which contains 51 data points in the redshift ranging from 0 to 2.36  \citep[see Table 1 in][]{2018CoTPh..70..445G}. These data are measured by the two methods. The first method is called the differential-age method, that is proposed to compare the ages of passively-evolving galaxies with similar metallicity, as cosmic chronometers, separated in a small redshift interval \citep{2002ApJ...573...37J} The second one is using the BAO measurement as a standard ruler along the radial direction \citep{Gaztanaga2009}.
%separated in a small redshift interval \citep{2002ApJ...573...37J} The second one is using the BAO measurement as a standard ruler along the radial direction \citep{Gaztanaga2009}.

The $\chi^2$ of the $H(z)$ data is given by
\begin{equation}
\chi^2_H = \sum_{i=1}^{N=51} \frac{\left[ H_{\rm obs}(z_i)-H_{\rm th}(z_i) \right]^2}{\sigma_H^2}.
\end{equation}
Here $H_{\rm obs}$ and $H_{\rm th}$ are the observational and theoretical Hubble parameters, respectively, and $\sigma_H$ is the error.

Finally, the joint $\chi^2$ of the three datasets is given by
\begin{equation}
\chi^2 = \chi^2_{\rm SN} + \chi^2_{\rm BAO} + \chi^2_H.
\end{equation}

By fitting these three datasets, we will constrain the mass of DE-like ULA filed $m_a$ and other parameters assuming $f_{\phi}=\phi_{\rm i}/M_{\rm pl}=1$ and $\le1$, respectively, in the flat and non-flat Universe.

\section{Constraint Results}\label{res}

In order to constrain the free parameters, the MCMC technique is adopted in this work. We make use of the public code $\tt Monte\ Python$\footnote{https://baudren.github.io/montepython.html} to perform the constraint, and the Metropolis-Hastings algorithm is employed to extract the chain points. Four cases in our model are explored, i.e. $f_{\phi}=1$ in flat and non-flat spaces, and $f_{\phi}\le1$  in flat and non-flat spaces. The flat priors are taken for the free parameters, and set as follow: $m_a/10^{-33} {\rm eV}\in(1,20)$, ${\rm log_{10}}f_{\phi}\in(-4,0)$, $h\in(0.5,1)$, $\Omega_m\in(0,1)$, and the SN Ia absolute magnitude $M\in(-30,-10)$. In each case, we generate 20 chains and totally obtain 1,000,000 chain points to illustrate the 1-D and 2-D probability distribution functions (PDFs) of the free parameters. 

\begin{figure*}[ht]
\begin{center}
\includegraphics[width=0.8\textwidth]{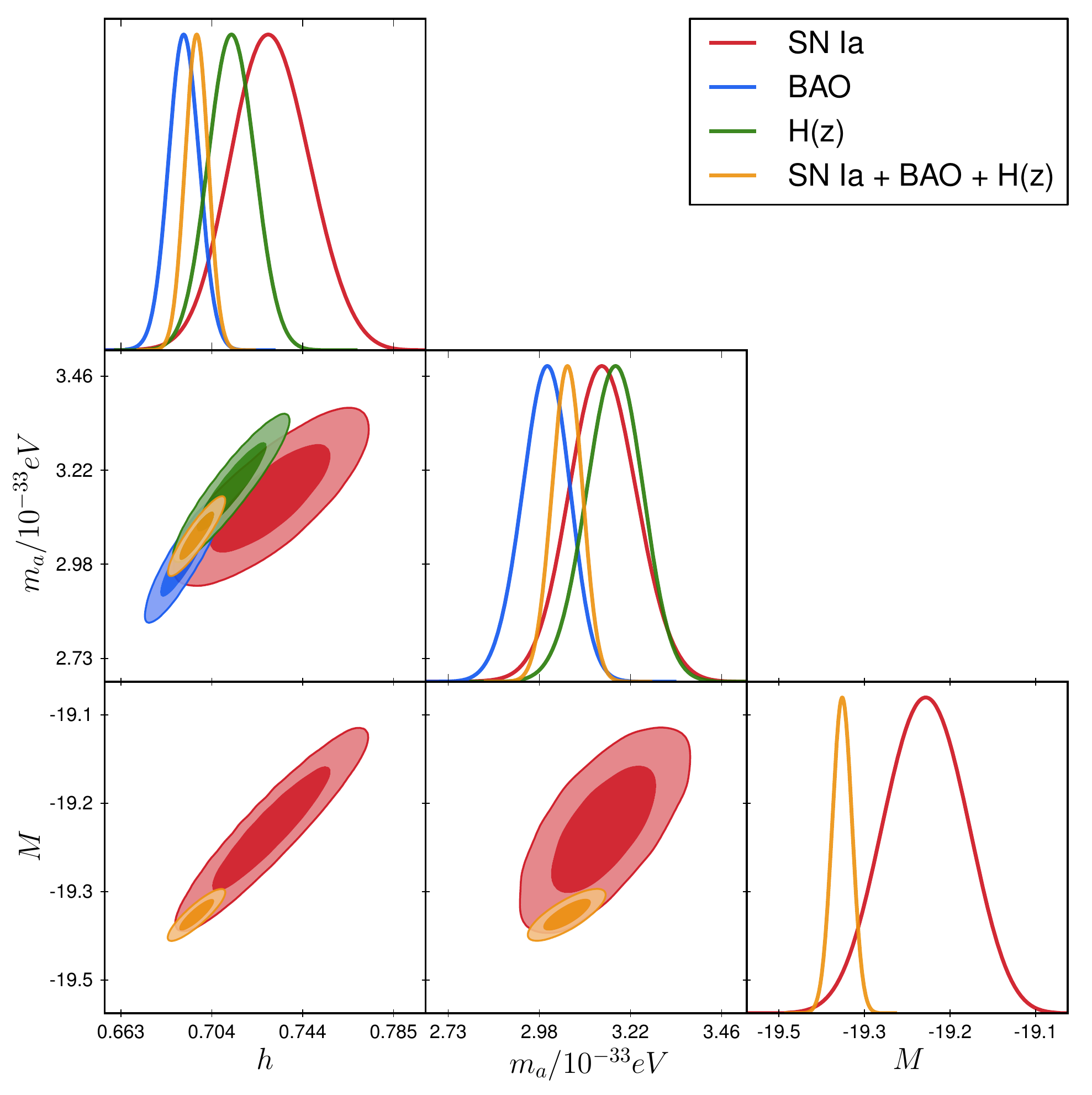}
\caption{\label{fig:f1flat} The constraint results of $m_a$, $h$, and $M$ for $f_{\phi}=1$ and flat case. The red, blue, green, and orange contours and curves are for the results from SN Ia, BAO, $H(z)$, and joint datasets, respectively. The 1-$\sigma$ (68.3\%), and 2-$\sigma$ (95.5\%) C.L. are shown for the contour maps.}
\end{center}
\end{figure*}

\begin{table*}[ht]\label{2d}
	\begin{center}
	\caption{\label{tab:f1flat}Best-fit values and 1-$\sigma$ errors of the parameters for $f_{\phi}=1$ and flat case.}
		\begin{tabular}{c|c|c|c|c}
			\hline\hline 
			Parameter&SN Ia& BAO &$H(z)$&SN Ia+BAO+$H(z)$\\
			\hline
			$h$&$0.730_{-0.018}^{+0.017}$&$0.691_{-0.007}^{+0.007}$&$0.712_{-0.011}^{+0.010}$&$0.697_{-0.005}^{+0.005}$\\
			$m_a/10^{-33}$ eV&$3.146_{-0.091}^{+0.091}$&$2.994_{-0.063}^{+0.067}$&$3.170_{-0.078}^{+0.075}$&$3.048_{-0.041}^{+0.042}$\\
			$M $&$-19.26_{-0.05}^{+0.05}$&$--$&$--$&$-19.37_{-0.01}^{+0.01}$\\
			$\chi^2_{\rm red}$&0.982&1.624&1.104&0.997\\
			\hline
		\end{tabular}
	\end{center}
\end{table*} % k=0

In Figure~\ref{fig:f1flat}, we show the contour maps and 1-D PDFs of $m_a$, $h$, and $M$ for $f_{\phi}=1$ and flat case. The best-fits and 1-$\sigma$ errors of the parameters for the SN Ia, BAO, $H(z)$, and joint datasets have been shown in Table~\ref{tab:f1flat}, respectively. The reduced chi-square $\chi^2_{\rm red}=\chi^2_{\rm min}/(N-M)$ is also shown, where $\chi^2_{\rm min}$ is the minimum $\chi^2$, and $N$ and $M$ are the numbers of data and free parameters, respectively. We find that our model can fit the data well, since $\chi^2_{\rm red}\sim1$ for the SN Ia, $H(z)$, and SN Ia+BAO+$H(z)$ data, and $\chi^2_{\rm red}\sim1.6$ for the BAO data. The best-fits of $h$ are around 0.7 for the BAO, $H(z)$, and SN Ia+BAO+$H(z)$ datasets from 0.69 to 0.71, and the result from SN Ia only is a bit high that $h\simeq0.73$. The results of $m_a$ are basically consistent in 1-$\sigma$ for the four datasets giving $m_a\simeq 3\times 10^{-33}$ eV. The nuisance parameter, the absolute magnitude $M$ shown in Eq.~(\ref{eq:mu_M}), in the SN Ia data is also considered in the fitting process, and the results from SN Ia only and SN Ia+BAO+$H(z)$ are in a good agreement.

\begin{figure*}[ht]
	\includegraphics[width=0.8\textwidth]{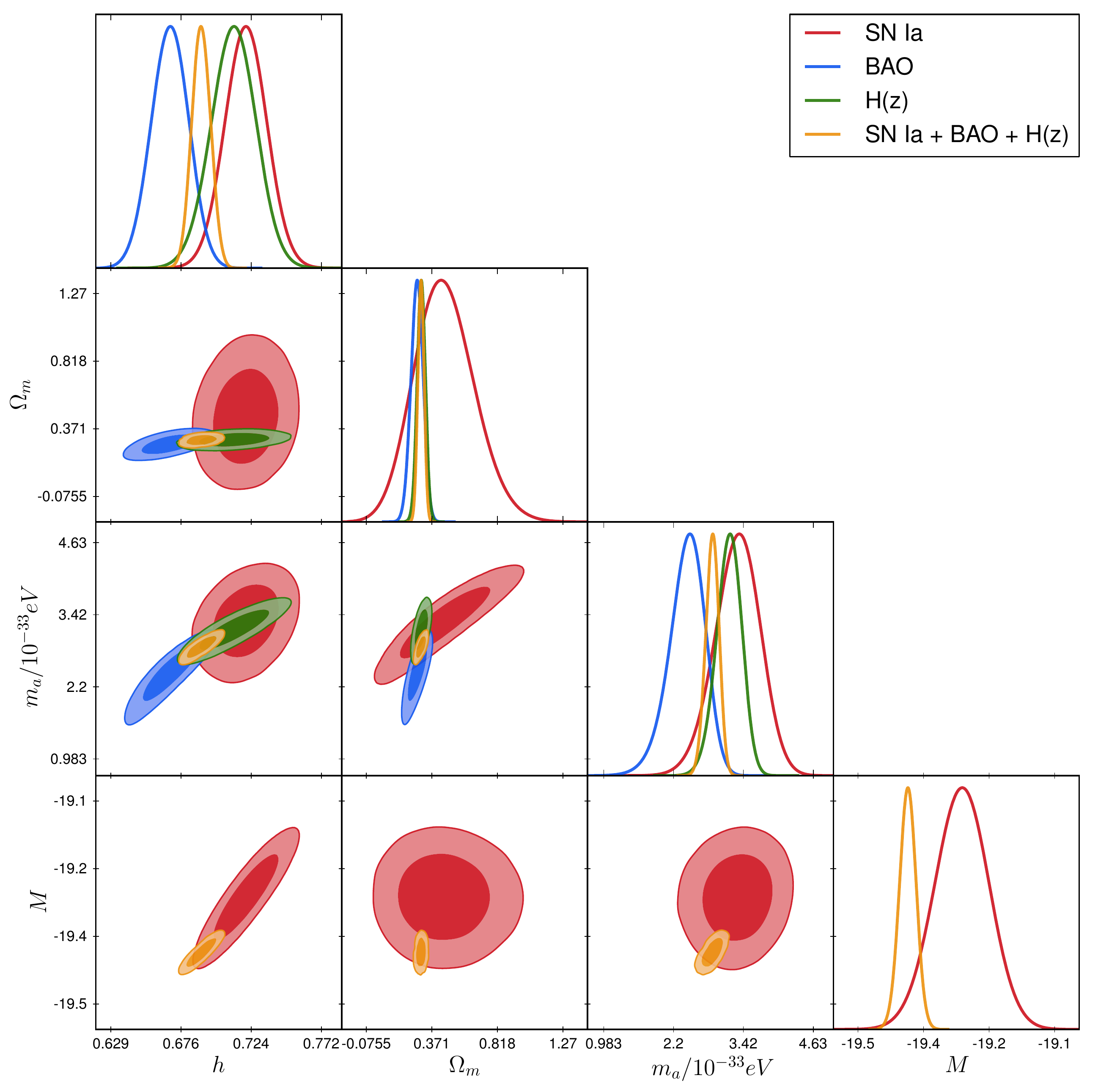}
	\caption{\label{fig:f1nonflat} Same as Figure~\ref{fig:f1flat}, but for the non-flat space. $\Omega_m$ is included in this case.}
\end{figure*}

\begin{table*}[ht]\label{2d}
	\begin{center}
	\caption{\label{tab:f1nonflat}Best-fit values and 1-$\sigma$ errors of the parameters for $f_{\phi}=1$ and non-flat case.}
		\begin{tabular}{c|c|c|c|c}
			\hline\hline
			Parameter&SN Ia&BAO &$ H(z) $&SN Ia+BAO+$H(z)$\\
			\hline
			$\Omega_m$&$0.384_{-0.221}^{+0.191} $&$0.274_{-0.041}^{+0.044}$& $0.300_{-0.029}^{+0.029}$&$0.295_{-0.021}^{+0.022}$\\
			$h$&$0.719_{-0.015}^{+0.015}$&$0.671_{-0.013}^{+0.013}$&$0.714_{-0.015}^{+0.016}$&$0.690_{-0.007}^{+0.006}$ \\
		    $m_a/10^{-33}$ eV&$3.332_{-0.382}^{+0.419} $&$2.485_{-0.281}^{+0.352}$&$3.191_{-0.210}^{+0.240}$&$2.868_{-0.110}^{+0.120}$ \\
			$\ M $&$-19.30_{-0.04}^{+0.04}$&$--$&$--$&$-19.42_{-0.01}^{+0.01}$\\			
			$\chi^2_{\rm red}$& 1.072&1.612 &1.129 &0.997\\
		\hline
		\end{tabular}
			\end{center}
\end{table*}

The fitting results of $\Omega_m$, $m_a$, $h$, and $M$ for $f_{\phi}=1$ and non-flat case are shown in Figure~\ref{fig:f1nonflat} and Table~\ref{tab:f1nonflat}. We find that the fitting is as good as the flat case, but the constraint results from SN Ia data is significantly looser than other datasets, which is due to the  combination effect of non-flat assumption and additional parameter $M$. We also notice that, although the best-fit of $\Omega_m$ from SN Ia only is as large as $\sim0.38$, it is consistent with the results from other dataset in 1-$\sigma$ with the best-fitting $\Omega_m\simeq0.3$. The SN Ia dataset gives obviously larger $h$ and $m_a$ compared to BAO dataset in this case, but basically the four datasets provide similar results as the flat case, that $h\sim0.7$ and $m_a\sim 3\times10^{-33}$ eV.

\begin{figure*}[ht]
	\begin{center}
		\includegraphics[width=0.8\textwidth]{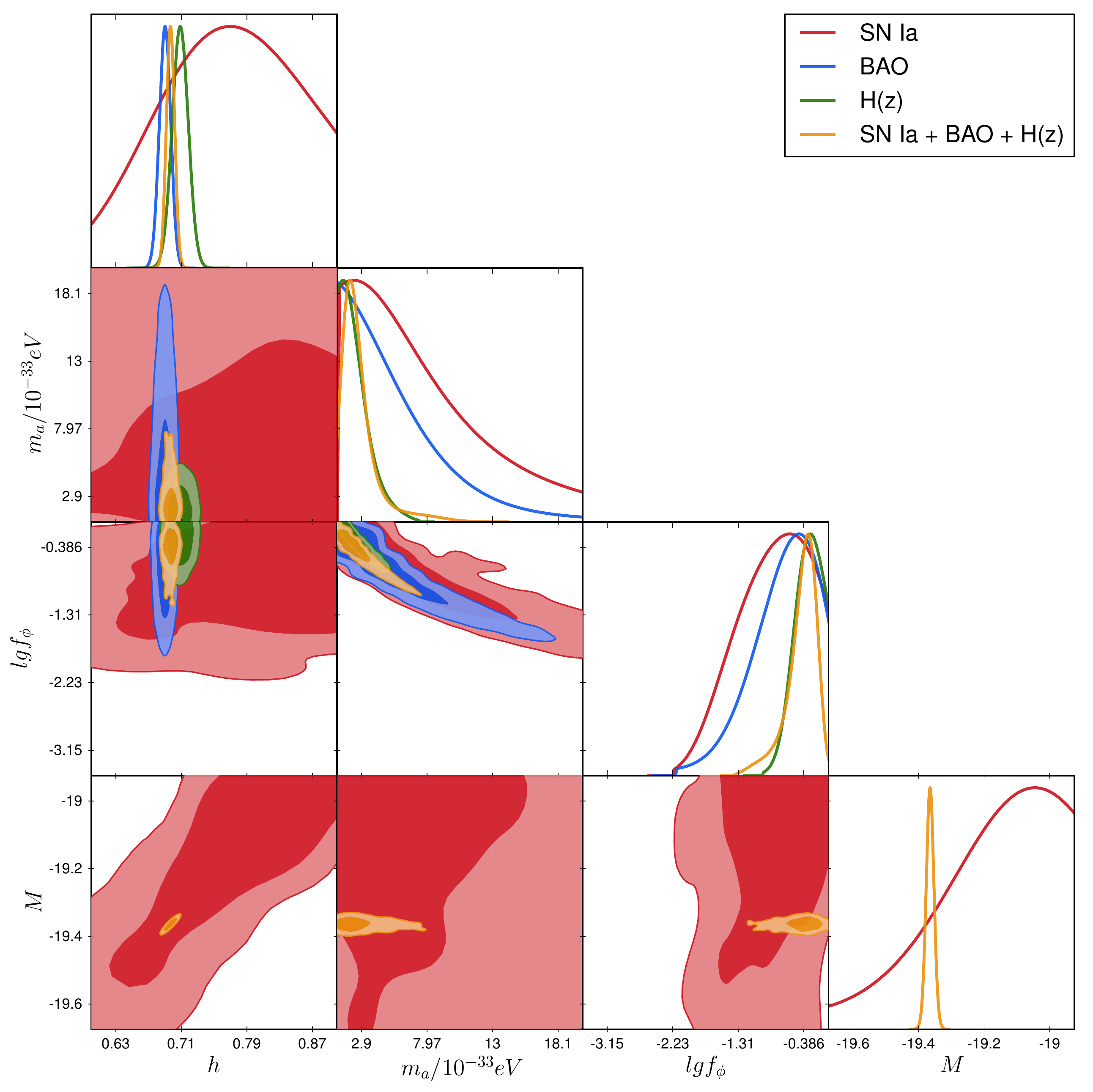}
		\caption{\label{fig:fl1flat} The constraint results of $m_a$, ${\log_{10}}f_{\phi}$, $h$, and $M$ for $f_{\phi}\le1$ and flat case. The red, blue, green, and orange contours and curves are for the results from SN Ia, BAO, $H(z)$, and joint datasets, respectively. The 1-$\sigma$ (68.3\%), and 2-$\sigma$ (95.5\%) C.L. are shown for the contour maps.}
\end{center}
\end{figure*}

\begin{table*}[ht]
	\begin{center}
	\caption{\label{tab:fl1flat} Best-fit values and 1-$\sigma$ errors of the parameters for $f_{\phi}\le1$ and flat case.}
		\begin{tabular}{c|c|c|c|c}
			\hline\hline 
			Parameter&SN Ia&BAO&$H(z)$&SN Ia+BAO+$H(z)$\\
			\hline
			$h$&$0.781_{-0.146}^{+0.122}$&$0.691_{-0.007}^{+0.007}$&$0.709_{-0.012}^{+0.009}$&$0.697_{-0.005}^{+0.005}$\\
			$m_a/10^{-33}$ eV&$2.734_{-2.734}^{+1.502}$&$1.250_{-1.210}^{+0.621}$&$1.785_{-1.193}^{+0.745}$&$2.403_{-1.520}^{+0.342}$\\
			${\rm log}_{10}f_\phi$&$-0.539_{-0.620}^{+0.539}$&$-0.433_{-0.360}^{+0.433}$&$-0.366_{-0.261}^{+0.310}$&$-0.379_{-0.095}^{+0.240}$\\
			$\ M $&$-19.08_{-0.29}^{+0.39} $&$--$&$--$&$-19.37_{-0.01}^{+0.01}$ \\
			$\chi^2_{\rm red}$& 1.011&1.382 &1.002&0.976\\
			\hline
		\end{tabular}
	\end{center}
\end{table*}

In Figure~\ref{fig:fl1flat} and Table~\ref{tab:fl1flat}, we show the constraint results of $m_a$, ${\log_{10}}f_{\phi}$, $h$, and $M$ for $f_{\phi}\le1$ and flat case. We find that the constraints on the parameters are generally looser than the $f_{\phi}=1$ case, especially for the SN Ia data (with additional nuisance parameter $M$), since there are strong degeneracies between $f_{\phi}$ and both $m_a$ and $h$ as shown in Eq.~(\ref{eq:Omega_a}). The SN Ia data cannot provide stringent constraint on $h$, which gives the best-fitting $h\simeq0.78$ with large error, but it is still consistent with the results from other datasets giving $h\simeq0.7$. The best-fits of $m_a$ are basically lower than the $f_{\phi}=1$ case, especially for the BAO data, that we have $m_a\simeq1-3\times10^{-33}$ eV with the lower limits of 1-$\sigma$ values close to zero for the four datasets. The best-fits of ${\rm log}_{10} f_{\phi}$ are around $-0.4$, and the upper limits of 1-$\sigma$ are consistent with 0 (i.e. $f_{\phi}\sim 1$) for the SN Ia, BAO, and $H(z)$ datasets. This means $\phi_{\rm i}\simeq M_{\rm pl}$ is a good assumption in this model.

\begin{figure*}[ht]
\begin{center}
	\includegraphics[width=0.8\textwidth]{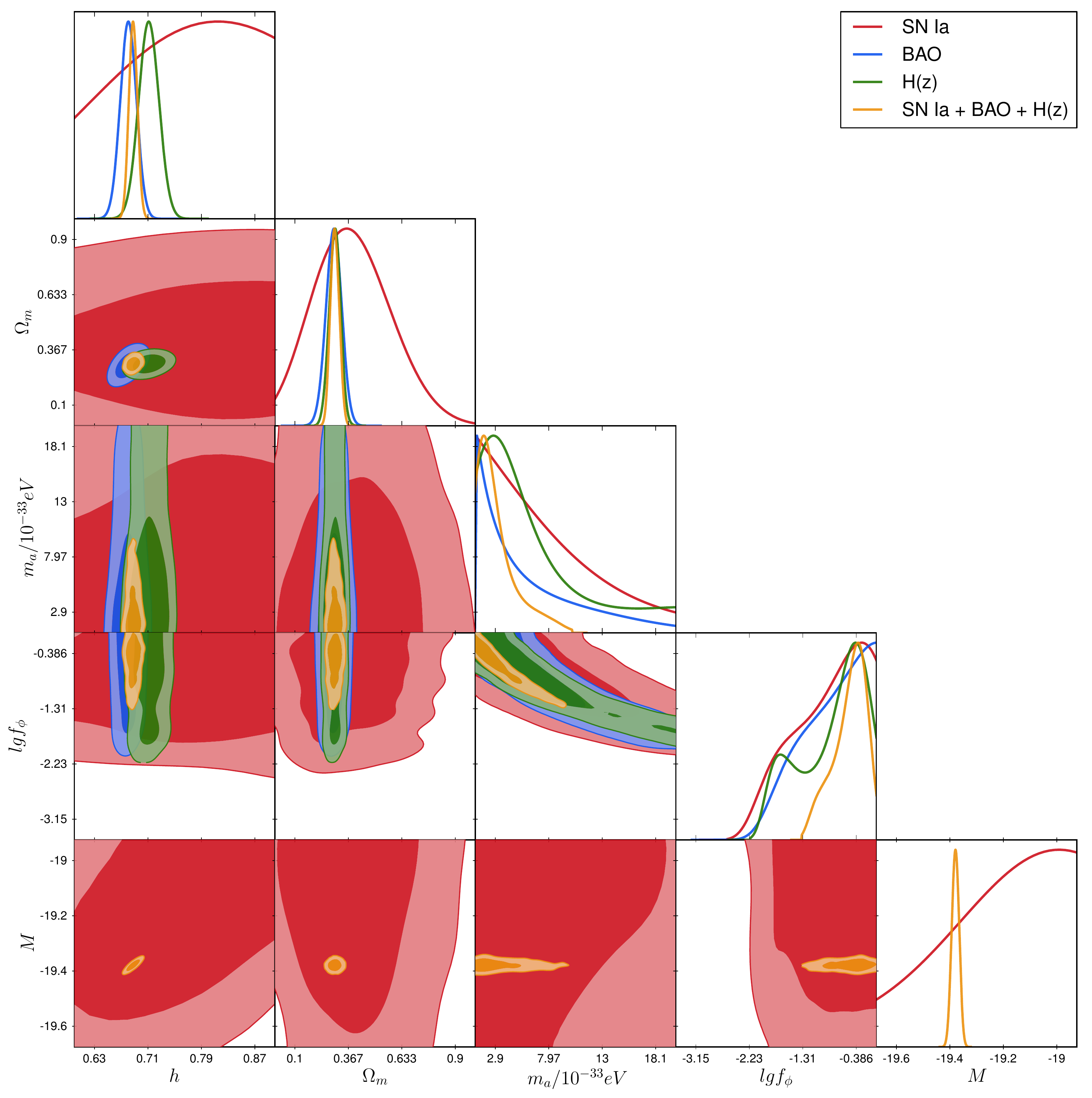}
	\caption{\label{fig:fl1nonflat}  Same as Figur~\ref{fig:fl1flat}, but for non-flat space. $\Omega_m$ is included in this case.}
\end{center}
\end{figure*}

\begin{table*}[ht]
	\begin{center}
	\caption{\label{tab:fl1nonflat} Best-fit values and 1-$\sigma$ errors of the parameters for $f_{\phi}\le1$ and non-flat case.}
		\begin{tabular}{c|c|c|c|c}
			\hline\hline
			Parameter&SN Ia&BAO &$ H(z) $&SN Ia+BAO+$H(z)$\\
			\hline
			$\Omega_m$&$0.367_{-0.049}^{+0.047} $&$0.308_{-0.042}^{+0.041}$&$0.310_{-0.028}^{+0.031}$&$0.305_{-0.022}^{+0.020}$\\
			$h$&$0.801_{-0.241}^{+0.130}$&$0.683_{-0.012}^{+0.012}$&$0.713_{-0.015}^{+0.016}$&$0.691_{-0.007}^{+0.006}$\\
			$m_a/10^{-33}$ eV&$1.024_{-1.024}^{+5.821} $&$1.009_{-1.009}^{+3.662}$&$2.791_{-2.791}^{+2.342}$&$2.121_{-1.072}^{+1.317}$ \\
			${\rm log}_{10}f_\phi$&$-0.344_{-0.331}^{+0.344}$&$-0.082_{-0.721}^{+0.082}$&$-0.367_{-0.811}^{+0.367}$&$-0.346_{-0.186}^{+0.245}$ \\
			$\ M $&$-19.03_{-0.49}^{+0.41} $&$--$&$--$&$-19.38_{-0.01}^{+0.01}$ \\
			$\chi^2_{\rm red}$& 1.009&1.281 & 1.103&0.981\\			
			\hline
		\end{tabular}
			\end{center}
\end{table*}

In Figure~\ref{fig:fl1nonflat} and Table~\ref{tab:fl1nonflat}, we show the fitting results for $f_{\phi}\le1$ and non-flat case. The constraint results are similar to the $f_{\phi}\le1$ and flat case, but have larger uncertainties since one more parameter $\Omega_m$ is included. Again, we find that the best-fits of $m_a$ are $1-3\times 10^{-33}$~eV, and lower limits of 1-$\sigma$ of $m_a$ are close to zero. Also, The upper limits of $f_{\phi}$ are consistent with (for SN Ia, BAO, and $H(z)$ only) or close to 1 (joint dataset).

\section{Summary}

In this work, we study the ULA field $\phi$ with mass around $10^{-33}$ eV that $m_a\sim H_0$, which means the ULA field still rolls slowly or just starts to enter the oscillation stage and can be treated as dark energy so far. We make use of the data from SN Ia, BAO, and $H(z)$ to constrain the properties of ULA field, and the MCMC technique is adopted to perform the data fitting process. The mass of the DE-like ULA field $m_a$, the ratio of initial field to the Planck mass $f_{\phi}$, the matter density parameter $\Omega_m$, reduced Hubble parameter $h$, and the SN Ia absolute magnitude $M$ are considered in the fitting process.

Four cases of the model are explored in this work, assuming the initial ULA field $\phi_{\rm i}=M_{\rm pl}$ and $\phi_{\rm i}\le M_{\rm pl}$ in flat and non-flat space, respectively. We find that the best-fits of $m_a$ are around $3\times 10^{-33}$ eV assuming $\phi_{\rm i}=M_{\rm pl}$ in either flat or non-flat space. When assuming $\phi_{\rm i}\le M_{\rm pl}$, the constraints become significantly looser, and we find the best-fits of $m_a$ become smaller, and the lower limits of 1-$\sigma$ are consistent with zero. Besides, the fitting results of ${\rm log}_{10}f_{\phi}$ are close to 0 for both flat and non-flat cases, which means the assumption of $\phi_{\rm i}=M_{\rm pl}$ is a good choice in this model. Besides the observational data adopted in this work, other datasets can be used to further improve the constraint results, such as the cosmic microwave background, weak lensing, galaxy cluster, and so on. We will investigate these constraint results in our future work.

\section{Acknownledgments}
JGK and YG acknowledges the support of NSFC-11822305, NSFC-11773031, NSFC-11633004, the Chinese Academy of Sciences (CAS) Strategic Priority Research Program XDA15020200, the NSFC-ISF joint research program No. 11761141012, and CAS Interdisciplinary Innovation Team.

\bibliographystyle{raa}
\bibliography{RAA-2019-0219}
 \end{document}